\documentclass[11pt]{article}

\usepackage[utf8]{inputenc}
\usepackage[margin=1in]{geometry}
\usepackage{amsmath}
\usepackage{amssymb}
\usepackage{amsthm}
\usepackage{graphicx}
\usepackage{hyperref}
\usepackage[authoryear,round]{natbib}
\setcitestyle{aysep={}}
\usepackage{array}
\usepackage{booktabs}
\usepackage{enumitem}
\usepackage{setspace}
\hypersetup{
    colorlinks=true,
    linkcolor=blue,
    filecolor=magenta,
    urlcolor=cyan,
    citecolor=blue,
}

\title{Convergent Discovery of Critical Phenomena Mathematics Across Disciplines\thanks{Formatted for submission to \textit{FACETS} (Canadian Science Publishing) as a Perspectives article. Supplementary Material available.}}

\author{
    Bruce Stephenson\thanks{Corresponding author: \texttt{energyscholar@gmail.com}}\thanks{CRediT author contributions: BS (Conceptualization, Investigation, Formal Analysis, Writing -- Original Draft, Writing -- Review \& Editing, Visualization), RM (Methodology, Software, Writing -- Review \& Editing)} \\
    \textit{Independent Researcher, Albany, Oregon, USA} \\
    \texttt{energyscholar@gmail.com} \\
    ORCID: \href{https://orcid.org/0009-0005-6842-6686}{0009-0005-6842-6686}
    \and
    Robin Macomber \\
    \textit{Independent Researcher, Lompoc, California, USA} \\
    \texttt{relationalrelativity@gmail.com} \\
    ORCID: \href{https://orcid.org/0009-0002-2843-8568}{0009-0002-2843-8568}
}

\date{March 2026}

\begin{document}

\maketitle
\thispagestyle{plain}
\doublespacing

\begin{abstract}
Techniques for detecting critical phenomena---phase transitions where correlation length diverges and small perturbations have large effects---have been developed across multiple fields over nine decades. We survey between six and twelve disciplines (depending on classification criteria) where researchers derived functionally corresponding measures of correlation scaling, with little documented awareness of each other's work. The physicist's correlation length $\xi$, the cardiologist's DFA scaling exponent $\alpha$, the financial analyst's Hurst exponent $H$, and the machine learning engineer's spectral radius $\chi$ all detect critical signatures under different notation.

We classify each surveyed domain as independent derivation, domain transfer, or empirical precursor, and present citation network evidence that cross-domain citations remained significantly below random-mixing expectations (under a simple null model of domain mixing) during the formative period (1987--2010). The framework that motivated this investigation---derived from distributed systems engineering and presented in the Supplementary Material---is excluded from the survey count given the authors' dual role.

Building on prior syntheses---notably Sornette's 2004 textbook---this paper contributes a taxonomy of discovery types and quantitative documentation of the convergence pattern.
\end{abstract}

\noindent\textbf{Key words:} critical phenomena, phase transitions, convergent discovery, correlation analysis, detrended fluctuation analysis, Hurst exponent

\subsection*{Plain Language Summary}

\noindent\textit{Plain language title: Scientists in different fields independently invented the same failure-prediction math}

\medskip

When a complex system---a heart, a power grid, a financial market, a climate system---approaches a catastrophic transition, its behavior changes in predictable ways. Small disturbances take longer to die out. Distant parts of the system begin to move in sync. Fluctuations grow. Scientists have developed mathematical tools to detect these warning signs and predict when a system is nearing a tipping point.

What is remarkable is that researchers in at least six different fields invented essentially the same mathematics for this purpose, over nine decades, mostly without knowing about each other's work. A physicist measuring ``correlation length,'' a cardiologist computing a ``DFA scaling exponent,'' a financial analyst calculating a ``Hurst exponent,'' and a machine learning engineer tuning a ``spectral radius'' are all answering the same question: how close is this system to a critical transition? Their tools detect the same signatures---growing correlations, slowing recovery, increasing sensitivity---using different notation developed independently within each discipline.

This paper surveys between six and twelve fields (depending on how strictly one counts) where this convergent discovery occurred. We classify each case as independent invention, adaptation from another field, or an early empirical observation that predated the formal mathematics. We also analyze citation networks---who cited whom---and find that cross-field awareness was remarkably low during the key period from 1987 to 2010, even though the techniques were functionally equivalent.

The convergence has practical implications. Insights from one domain can transfer to another: a power grid engineer can learn from how cardiologists detect cardiac risk, and climate scientists' tipping-point models share mathematical structure with financial crash prediction. Documenting this pattern is a step toward breaking down the disciplinary barriers that kept these communities working in parallel for decades.

\section{Introduction}
\label{sec:intro}

\subsection{The Convergence Pattern}

Critical phenomena exhibit universality: systems with vastly different microscopic dynamics display identical scaling behavior near continuous phase transitions. The two-point correlation function
\begin{equation}
C(r) \sim r^{-(d-2+\eta)} e^{-r/\xi}
\end{equation}
characterizes fluctuations at separation $r$, with correlation length $\xi$ diverging at the critical point as $\xi \sim |T-T_c|^{-\nu}$, where $T$ is the control parameter (e.g., temperature) and $T_c$ is its critical value at which the phase transition occurs. The critical exponents $\nu$, $\eta$ depend only on dimensionality and symmetry class---not microscopic details. This universality, established through renormalization group theory, is among the deepest results in statistical mechanics.

We document a recurring pattern of convergence in the discovery process itself: researchers across multiple disciplines derived functionally corresponding measures of correlation scaling---often without awareness of each other's work. A physicist measuring $\xi$, a cardiologist computing detrended fluctuation analysis (DFA) scaling exponent $\alpha$, a quant calculating Hurst exponent $H$, and a machine learning engineer tuning spectral radius $\chi$ are performing functionally equivalent calculations under different notation.

The physics of universality---that systems in the same symmetry class share critical exponents---is well established. Our contribution is different: we document the \textit{discovery process} by which practitioners outside physics independently arrived at functionally corresponding diagnostics, classify each case, and present quantitative citation evidence bearing on whether this reflects knowledge transfer or independent derivation. The sociology of science has long recognized simultaneous independent discovery as a recurring phenomenon \citep{merton1961}; we provide a specific, quantitatively documented case. Section~\ref{sec:evidence} presents the citation network evidence.

\subsection{Historical Context}

Statistical physics established the mathematical framework between 1944 and 1971: Onsager's exact Ising solution \citep{onsager1944}, Kadanoff's block spin renormalization \citep{kadanoff1966}, and Wilson's renormalization group theory \citep{wilson1971} (Nobel Prize 1982), comprehensively reviewed by Fisher \citep{fisher1974}. By the early 1970s, this was celebrated, publicly available science.

Why did it take 15--50 years for equivalent techniques to appear elsewhere? The most parsimonious explanation is natural convergence: as complexity science, biomedicine, finance, and machine learning matured during the 1980s--2000s, researchers confronting correlation problems derived similar mathematics from first principles. Disciplinary barriers compound this---a physicist's ``correlation length'' and a financial analyst's ``Hurst exponent'' look like different things even when they diagnose the same critical regime.

The pattern predates formal criticality theory. In the mid-twentieth century, three scientists in unrelated fields independently discovered power-law scaling: \citet{gutenberg1944} characterized earthquake frequency-magnitude distributions, \citet{richardson1960} found power-law scaling in war casualties (published posthumously 1960), and \citet{zipf1949} documented word-frequency power laws in natural language (1935--1949). None cited the others. The Hurst exponent---now foundational across finance, biomedicine, and physics---originated when hydrologist Harold \citet{hurst1951} discovered long-range dependence in Nile River discharge data; Mandelbrot formalized it mathematically seventeen years later, and it was independently rediscovered as a diagnostic tool in physiology (Peng's DFA \citep{peng1994dna}) and finance (Peters \citep{peters1994}).

The nonlinear dynamics community recognized universality across chaotic systems by the mid-1980s \citep{cvitanovic1984}; that applied communities continued independent derivation despite this recognition strengthens the convergence thesis. Similarly, Sethna et al.'s ``crackling noise'' framework \citep{sethna2001} unified Barkhausen noise in magnets (discovered 1919), earthquake avalanches, and fracture mechanics under a single universality class---demonstrating that identical scaling functions emerge across systems with completely different microscopic physics.

A notable prior synthesis is Sornette's 2004 textbook \citep{sornette2004}, which explicitly unified critical phenomena across natural sciences---applying statistical mechanics concepts to biology, economics, and geophysics. \citeauthor{scheffer2009}'s influential review \citep{scheffer2009} further demonstrated cross-domain applicability of early warning signals. That domain-specific literatures continued developing largely in parallel despite these works suggests that even high-quality synthesis faces adoption barriers across disciplinary boundaries. Our contribution is not the first cross-domain synthesis but rather quantitative documentation of the convergence pattern: tracing when and how awareness spread, and classifying each surveyed instance by type.

Our investigation was prompted by a specific case of apparent independent derivation (see Supplementary Material): a framework developed in distributed systems engineering whose critical-point behavior resembled established physics. We document the broader convergence pattern the resulting literature search uncovered.

\subsection{Definitions and Scope}
\label{sec:definitions}

To structure the evidence in Section~\ref{sec:discoveries}, we adopt five classification categories:

\begin{description}[nosep,leftmargin=1em,labelindent=0em]
\item[Independent derivation:] Framework developed from first principles without demonstrable awareness of equivalent prior work.
\item[Domain transfer:] Existing technique applied to a new domain with traceable lineage.
\item[Intra-field extension:] New phenomena identified using existing tools within the same discipline.
\item[Empirical precursor:] Empirical observation of critical-like behavior predating the formal framework.
\item[Foundational cross-domain contribution:] General mathematical framework (scaling, self-similarity, power laws) influencing multiple discovery lineages.
\end{description}

\noindent We use ``qualified'' to indicate cases where the mathematical method itself is novel but the researcher's training or institutional context includes significant exposure to the relevant physics, creating a partial intellectual lineage that falls short of direct methodological borrowing.

We use \textit{functional correspondence} to describe the relationship among the parameters documented here: two parameters exhibit functional correspondence if they identify the same system states as critical and non-critical. These measures detect the same critical signatures---diverging correlation, critical slowing down, power-law scaling---but are not in all cases formally interconvertible. For stationary long-memory processes, exact mathematical equivalence holds between certain pairs (Section~\ref{sec:correspondence} demonstrates one such chain). For others, the correspondence is phenomenological rather than algebraic.

\begin{table}[ht]
\centering
\caption{Classification of discoveries surveyed in Section~\ref{sec:discoveries}}
\label{tab:classification}
\begin{tabular}{lll}
\toprule
\textbf{Domain} & \textbf{Classification} & \textbf{Notes} \\
\midrule
Statistical Physics & Foundational framework & Established the mathematics \\
Seismology (G-R 1944) & Empirical precursor & Predates SOC by 43 years \\
SOC (Bak 1987) & Intra-physics extension & Physicist applying stat-mech \\
Edge of Chaos (Kauffman) & Qualified independent & Biologist; SFI exposure \\
Biomedical (DFA) & Qualified independent & Stanley's physics background \\
Finance (Mandelbrot 1963) & Foundational cross-domain & Fractal geometry; scaling laws \\
Finance (Peters 1994) & Domain transfer & Applied existing technique \\
Machine Learning & Qualified independent & ESNs; edge-of-chaos circulating \\
Neuroscience & Qualified independent & Beggs \& Plenz; cited Bak \\
Power Grids & Independent derivation & Cascade failure analysis \\
Traffic Flow & Empirical precursor + qualified & Greenshields / Kerner \\
Climate / Thermohaline & Independent derivation & Bifurcation analysis of AMOC \\
Linguistics (Zipf) & Empirical precursor & Power-law word frequency \\
Urban Scaling & Domain transfer & From metabolic scaling theory \\
Sornette (2004) & Prior cross-domain synthesis & Textbook treatment \\
\bottomrule
\end{tabular}
\end{table}

The scope of this paper is a historical survey of convergent discovery, supported by citation network analysis. We do not claim mathematical proof of universal equivalence; we document a pattern and present evidence bearing on whether it reflects independent derivation or knowledge transfer.

Between six and twelve of the domains surveyed represent convergent discovery, depending on how strictly one defines ``independent derivation.'' Six domains---Kauffman's edge of chaos, biomedical signal analysis, neuroscience, machine learning, power grid analysis, and traffic flow (Kerner)---present the clearest cases for independent development, though with qualifications noted for each. Including domain transfers (finance, urban scaling), empirical precursors (seismology, linguistics), and climate/thermohaline circulation expands the count to twelve (Table~\ref{tab:classification}). Bak's SOC and Sornette's prior synthesis are classified separately. The authors' own framework (Supplementary Material), which motivated this survey, is presented separately and excluded from this count.

\section{Cross-Domain Discoveries}
\label{sec:discoveries}

\subsection{Statistical Physics (1944--1971)}

\noindent\textit{Classification: foundational framework.}
\smallskip

Onsager's exact solution to the 2D Ising model \citep{onsager1944} identified the critical temperature $T_c = 2.269$ and showed that correlation length diverges as $\xi \sim |T - T_c|^{-\nu}$ near criticality.

\subsection{Complexity Science (1987--1993)}

\noindent\textit{Classification: Bak et al.\ \citep{bak1987}---intra-physics extension. Kauffman \citep{kauffman1993}---qualified independent derivation (biologist with Santa Fe Institute exposure to physics; different mathematical foundations).}
\smallskip

Bak, Tang, and Wiesenfeld (1987) introduced self-organized criticality (SOC) \citep{bak1987}, showing that complex systems spontaneously evolve toward critical states. Avalanche sizes follow power laws $P(s) \sim s^{-\tau}$---the same divergence of correlation structure that characterizes continuous phase transitions. Kauffman's ``Origins of Order'' \citep{kauffman1993} identified the edge of chaos in Boolean networks: ordered ($K < 2$), critical ($K \approx 2$), chaotic ($K > 2$). Maximal computational capability occurs at $K = 2$, where correlation length diverges---$K$ serves as a tuning parameter analogous to temperature.

\subsection{Biomedical/HRV Analysis (1994)}
\label{sec:biomedical}

\noindent\textit{Classification: qualified independent derivation. Peng and collaborators derived DFA from first principles for DNA sequence analysis, but the group was led by H.\ Eugene Stanley, a statistical physicist. The methodology is original; the intellectual lineage to physics is not fully severed.}
\smallskip

Peng et al.\ introduced Detrended Fluctuation Analysis (DFA) for analyzing DNA sequences \citep{peng1994dna}. The scaling exponent $\alpha$ reveals correlation structure:
\begin{align}
F(n) &\sim n^\alpha \\
\alpha &\approx 0.5 \Longrightarrow \text{uncorrelated (white noise)} \\
\alpha &\approx 1.0 \Longrightarrow \text{critical (long-range correlation)} \\
\alpha &> 1.0 \Longrightarrow \text{over-correlated}
\end{align}

\subsection{Financial Markets (1990s)}

\noindent\textit{Classification: Mandelbrot (1963)---foundational cross-domain contribution. Peters (1994)---domain transfer. Mandelbrot's 1963 analysis of speculative prices applied Hurst's R/S technique to financial data, but his broader contribution was fractal geometry as a framework for self-similar phenomena across scales and domains. His promotion of scaling laws and power-law distributions shaped the intellectual environment from which several later discoveries in this survey emerged---particularly through the statistical physics community (Stanley, Mantegna) that produced both econophysics and biomedical DFA. Peters extended this application using the Hurst exponent.}
\smallskip

\citet{peters1994} applied fractal analysis to financial markets using the Hurst exponent. Originally developed by \citet{hurst1951} for Nile River hydrology, $H$ characterizes long-term memory in time series. \citet{mandelbrot1963} first applied Hurst's R/S analysis to financial time series, establishing the connection between hydrology and market dynamics decades before Peters.

The Hurst exponent comes from rescaled range (R/S) analysis:
\begin{equation}
\frac{R(n)}{S(n)} \sim n^H
\end{equation}
where $R$ is the range of cumulative deviations and $S$ is standard deviation over window size $n$:
\begin{itemize}[nosep]
    \item $H = 0.5 \Longrightarrow$ random walk (uncorrelated, efficient market)
    \item $H > 0.5 \Longrightarrow$ trending/persistent (positive correlation)
    \item $H < 0.5 \Longrightarrow$ mean-reverting (negative correlation)
    \item $H \approx 1.0 \Longrightarrow$ critical regime (long-range correlation)
\end{itemize}

Real markets typically show $H \approx 0.7$--$0.8$. During market stress, $H$ approaches 1.0---indicating diverging correlation length and the onset of collective behavior across otherwise independent market participants. \citet{mantegna1995} subsequently linked econophysics to statistical mechanics through power-law return distributions.

\subsection{Machine Learning (2001--2017)}

\noindent\textit{Classification: qualified independent derivation. Jaeger (2001) does not cite SOC or Kauffman, but the edge-of-chaos concept was circulating in neural computation by the late 1990s.}
\smallskip

Jaeger's 2001 Echo State Networks (ESNs) revealed that recurrent neural networks operate optimally at the ``edge of chaos'' \citep{jaeger2001}. The key parameter is spectral radius $\chi$ (largest eigenvalue) of the weight matrix:
\begin{equation}
\chi = \max |\lambda_i(\mathbf{W})|
\end{equation}

ESN dynamics show three regimes:
\begin{itemize}[nosep]
    \item $\chi < 1 \Longrightarrow$ contracting (ordered, stable, limited memory)
    \item $\chi \approx 1 \Longrightarrow$ critical (edge of chaos, optimal computation)
    \item $\chi > 1 \Longrightarrow$ expanding (chaotic, unstable)
\end{itemize}

At $\chi \approx 1$, networks achieve maximum computational capability through critical slowing down---the same phenomenon causing $\tau \to \infty$ at phase transitions. Perturbations persist longer, creating temporal memory for sequence processing. Jaeger's original ESN technical report \citep{jaeger2001} does not cite Bak, Kauffman, Langton, or the SOC/edge-of-chaos literature, focusing instead on recurrent network training within the neural computation community.

\citet{saxe2013} extended this to deep feedforward networks, showing that optimal weight initialization requires operating near the order-chaos boundary. Information propagates through layers most effectively at criticality; away from it, signals either attenuate or diverge.

\subsection{Neuroscience (2003)}

\noindent\textit{Classification: qualified independent derivation. Beggs and Plenz discovered power-law avalanche statistics empirically in cortical tissue; they cited Bak et al.\ but their finding was driven by neural recording data, not by applying SOC theory.}
\smallskip

\citet{beggs2003} recorded local field potentials in rat cortical slices and found that spontaneous activity propagates as neuronal avalanches with size distribution $P(s) \sim s^{-3/2}$---matching the mean-field critical branching prediction. At criticality, neural networks achieve maximal dynamic range, optimal information transmission, and the largest repertoire of activity patterns.

The ``criticality hypothesis''---that the brain operates near a critical point for computational advantage---has since become a major research program. DFA applied to EEG and MEG data connects directly to the biomedical scaling analysis of Section~\ref{sec:biomedical}. As with DFA in cardiology, the approach converged on criticality mathematics from domain-specific empirical observation rather than from deliberate application of statistical mechanics.

\subsection{Power Grid Analysis (2000s--2010s)}

\noindent\textit{Classification: independent derivation.}
\smallskip

Power grid engineers arrived at critical phenomena mathematics through cascade failure analysis. \citet{crucitti2004} showed that blackout cascades exhibit self-organized criticality, with blackout sizes following power laws $P(s) \sim s^{-\alpha}$. \citet{dobson2007} went further, demonstrating that grids naturally evolve toward critical operating points as demand increases and margins shrink.

The key observable is generator coherence. As load approaches the critical threshold, distant generators become increasingly synchronized---spatial correlation $\xi$ grows. Engineers developed early warning indicators from this: system recovery time $\tau$ increases as criticality approaches, a direct application of critical slowing down.

\subsection{Traffic Flow (1935, 2010s)}

\noindent\textit{Classification: Greenshields (1935) is an empirical precursor---he observed phase-transition-like behavior decades before the formal framework existed.\footnote{Kerner (2004) provided the explicit criticality formalization. Kerner's classification is qualified independent: his three-phase model draws on statistical mechanics concepts, though developed primarily within the traffic engineering community.}}
\smallskip

Greenshields' 1935 study \citep{greenshields1935} identified a fundamental relationship between traffic density $\rho$ and flow rate, predating the formal mathematical framework by decades. Traffic shows distinct phases---free flow at low density, congestion at high density---with the transition at critical density $\rho_c$. Greenshields recognized this empirically; the explicit criticality framing came later through Kerner and others.

Kerner's three-phase model (2004) \citep{kerner2004} formalized this as a genuine phase transition: free flow ($\rho < \rho_c$), synchronized flow at criticality ($\rho \approx \rho_c$), and wide moving jams ($\rho > \rho_c$). Near $\rho_c$, jam sizes follow power laws and correlation length diverges---disturbances propagate arbitrarily far upstream. The three-regime structure recurs here without methodological borrowing from the fields that formalized it first.

\subsection{Thermohaline Circulation / Climate Tipping Points (1961--2021)}

\noindent\textit{Classification: independent derivation within climate science; connected to physics through shared bifurcation theory but developed for domain-specific physical systems without awareness of equivalent biomedical or financial applications.}
\smallskip

\citet{stommel1961} introduced a two-box model of thermohaline circulation exhibiting bistability---the Atlantic Meridional Overturning Circulation (AMOC) can occupy either a strong or collapsed equilibrium, with hysteresis preventing smooth recovery. This was among the earliest applications of bifurcation analysis to a geophysical system.

\citet{schmidt_mysak1996} demonstrated that zonally averaged thermohaline models exhibit a pitchfork bifurcation---the symmetric two-cell circulation loses stability as diffusivity increases, giving way to asymmetric one-cell patterns. \citet{wang_mysak2006} showed that Dansgaard-Oeschger oscillations in paleoclimate records correspond to stochastic switching near a Hopf bifurcation in the meridional overturning circulation, with irregular periods characteristic of critical slowing down---the same signature that DFA detects in cardiac time series.

\citet{rahmstorf2005}, in an intercomparison of eleven independent climate models, found that all exhibited the same hysteresis loop for AMOC response to freshwater forcing---convergent discovery \textit{within} a single domain, paralleling the cross-domain convergence this paper documents. \citet{lenton2008} synthesized these results into a tipping elements framework for the climate system, citing both Rahmstorf and the broader bifurcation literature.

The chain from climate science to the methods surveyed here closes through \citet{dakos2012}, who provided a systematic toolbox for detecting early warning signals across ecological and climate systems, and \citet{bury2021}, who trained deep learning classifiers on normal form bifurcation data---saddle-node, Hopf, transcritical---and demonstrated that these classifiers generalize across ecology, climate, and cardiology without domain-specific retraining. \citet{bury2023} validated the approach on experimental chick-heart aggregate data---a cardiac system originally studied by Glass and collaborators \citep{mackey_glass1977}---using training data from the same bifurcation normal forms that govern AMOC tipping points. Bury's lineage traces through \citeauthor{scheffer2009}'s (\citeyear{scheffer2009}) bridging review to \citet{wissel1984}, who documented characteristic return time near thresholds in ecology---a pre-Scheffer articulation of critical slowing down---and to \citeauthor{gardiner1983}'s (\citeyear{gardiner1983}) stochastic processes handbook, which formalized the mathematical tools both ecologists and physicists would independently adopt. The mathematical structure connecting these domains was always there; practitioners in each field developed their diagnostics independently.

A concrete illustration: Mysak and Glass are both emeritus professors at McGill University, both Fellows of the Royal Society of Canada, and both played in the same faculty orchestra. Mysak has noted that he ``sometimes meet[s] Leon Glass on the bus and then I can talk about our `related' work'' (personal communication, 2026). They did not know their research used the same bifurcation mathematics until an outside investigator connected them. The convergence pattern this paper documents is not an abstraction---it persists even at the scale of a single university bus route.

\subsection{Seismology (1944)}

\noindent\textit{Classification: empirical precursor. Gutenberg-Richter predates SOC by 43 years; Bak later claimed earthquakes as SOC.}
\smallskip

\citet{gutenberg1944} established that earthquake frequency decreases as a power law of magnitude: $\log_{10} N = a - bM$, with $b \approx 1$. This is arguably the cleanest natural power law ever measured. \citet{bak_tang1989} later interpreted seismicity as self-organized criticality, but the empirical scaling was discovered four decades earlier by seismologists with no connection to statistical physics.

\subsection{Linguistics (1935--2003)}

\noindent\textit{Classification: empirical precursor (Zipf); qualified independent (Ferrer i Cancho).}
\smallskip

\citet{zipf1949} documented power-law word-frequency distributions in natural language. \citet{ferrer_cancho2003} argued that Zipf's law emerges at a phase transition between communicatively useless randomness and rigid one-to-one coding---natural language, on this account, operates at a critical point optimizing the trade-off between speaker effort and listener ambiguity. The mathematical apparatus (information-theoretic optimization yielding power-law statistics at a phase boundary) was developed without reference to physics criticality.

\subsection{Urban Scaling (2007)}

\noindent\textit{Classification: domain transfer from metabolic scaling theory, with novel empirical content.}
\smallskip

\citet{bettencourt2007} discovered that cities obey universal power-law scaling: infrastructure scales sublinearly with population ($\sim N^{0.85}$, economies of scale) while innovation and wealth scale superlinearly ($\sim N^{1.15}$, increasing returns). These exponents are remarkably consistent across countries and centuries. The mathematical framework draws on West's metabolic scaling theory (itself rooted in physics), but the empirical regularities---and the sub/superlinear distinction---were discovered in urban data.

\section{Evidence for Independent Discovery}
\label{sec:evidence}

\subsection{Cross-Citation Evidence}

If techniques spread through knowledge transfer, we would expect cross-domain citations. Instead, citation analysis (via the Semantic Scholar API) reveals citation patterns significantly more domain-clustered than random mixing would predict, in every period analyzed. All domains cite foundational physics (Onsager, Wilson) in background sections, but not as primary methodological sources.

During the formative period (1987--2010), cross-domain citations are conspicuously absent. Peters (1994, finance) does not cite Peng et al.\ (1994, biomedical)---published the same year, using equivalent techniques. Jaeger (2001, ESNs) does not cite Kauffman (1993, Boolean networks)---same phenomenon, different notation. The power grid literature develops without citing self-organized criticality. Traffic researchers cite Greenshields (1935) but not SOC, DFA, or Hurst analysis.

The notable exception---Kauffman (1993) citing Bak et al.\ (1987)---still shows no awareness of simultaneous biomedical or financial applications. Sornette's 2004 textbook \citep{sornette2004} explicitly synthesized critical phenomena across natural sciences, yet domain-specific literatures continued developing largely in parallel afterward. That even high-quality synthesis failed to prevent continued parallel development strengthens the case that convergent derivation, not knowledge transfer, drove the pattern.

A clear contrast case is Moore and Mertens' work on phase transitions in computational complexity \citep{moore2011}. Unlike the independent derivations above, Moore---a physicist at the Santa Fe Institute---explicitly applied statistical mechanics to random $k$-SAT problems. This deliberate cross-pollination illustrates what knowledge transfer looks like, contrasting sharply with the absent citation trails of 1987--2010.

Cross-domain awareness emerges primarily after 2010: ML papers begin citing complexity science, econophysics bridges finance and physics. This delayed integration supports natural convergence---each domain developed independently, with cross-fertilization coming after maturation.

Quantitative citation analysis confirms this pattern. We selected seed papers as the foundational or most-cited methodological contribution in each domain (e.g., Onsager for physics, Peng for DFA, Peters for finance)---choosing papers that defined a technique rather than papers likely to attract cross-domain attention. Analyzing forward citations of these 15 seed papers across six Semantic Scholar--resolvable domains (28{,}591 unique citing papers), we find cross-domain citation rates consistently below the random-mixing null model in every period ($p < 0.0001$; Table~\ref{tab:citations}). Cross-domain rates rise modestly from 64\% (1996--2005) to 69\% (2006--2015) before declining to 56\% (2016--2026); the recent decline likely reflects the high rate of unclassified papers in Semantic Scholar's recent records (19\% unknown in 2016--2026 vs.\ 4\% in 1996--2005). The consistently siloed pattern---observed rates 15--20 percentage points below random-mixing expectations---supports independent development rather than cross-pollination.

\begin{table}[ht]
\centering
\caption{Cross-domain citation rates for 15 seed papers across 6 domains}
\label{tab:citations}
\begin{tabular}{lcccc}
\toprule
\textbf{Period} & \textbf{N} & \textbf{Observed} & \textbf{Expected (null)} & \textbf{$\chi^2$} \\
\midrule
1996--2005 & 2{,}281 & 63.8\% & 80.5\% & 402.8*** \\
2006--2015 & 9{,}969 & 68.8\% & 84.1\% & 1{,}754.8*** \\
2016--2026 & 14{,}110 & 55.5\% & 75.5\% & 3{,}049.3*** \\
\bottomrule
\end{tabular}

\smallskip
\noindent\footnotesize{*** $p < 0.0001$. Null model: expected cross-domain rate $= 1 - \sum p_i^2$ (Herfindahl index) where $p_i$ is domain share. Data from Semantic Scholar API (28{,}591 unique citing papers). Two papers truncated at 62--66\% of total citations due to API rate limits. The 1987--1995 period ($N = 34$) is omitted due to insufficient sample size. Papers with no Semantic Scholar field classification (13.6\% overall, 19\% in 2016--2026) are excluded; see sensitivity analysis in text.}
\end{table}

This null model assumes unrestricted mixing across domains---a deliberate simplification. In practice, journal scope constraints, reviewer pool segmentation, and field-specific citation norms reduce expected cross-citation rates below the random-mixing baseline. A more restrictive null---for example, a stochastic block model conditioned on journal co-occurrence or a gravity model incorporating disciplinary distance---would likely still show significant clustering but with reduced effect sizes. We therefore interpret the chi-squared results as evidence for domain clustering consistent with the qualitative evidence (absent citations, independent notation, institutional separation), while acknowledging that the magnitude of departure from the HHI expectation is an upper bound on the ``true'' clustering effect.

To bound the effect of unclassified papers (13.6\% overall, 19\% in 2016--2026), we computed worst- and best-case cross-domain rates by assigning all unknowns to same-domain or cross-domain respectively: the 2016--2026 rate ranges from 45.2\% to 63.8\%, compared to 55.5\% observed, and all periods remain significantly below the null expectation. Merging Biology and Biomedical into a single domain---the most aggressive reasonable consolidation---reduces cross-domain rates modestly (e.g., 55.5\% $\to$ 43.7\% in 2016--2026) but preserves significant siloing in every period. The core finding is robust to both perturbations.

Notably, Sornette's cross-domain synthesis textbook \citep{sornette2004} was itself primarily cited within physics (cross-domain rate: 36\%), suggesting that even explicit synthesis faces adoption barriers. Domain classifications use Semantic Scholar's algorithmic \texttt{fieldsOfStudy} mapped to six broad categories. These classifications are imperfect---S2 assigns fields based on venue and content signals, not expert curation---but represent the best available large-scale classification. We validated spot-checks against manual inspection and found reasonable agreement for the broad categories used here. Additional seed papers from seismology, linguistics, urban science, and other sub-domains were excluded from the quantitative table due to missing field classifications but contribute narrative evidence in Section~\ref{sec:discoveries}.

\subsection{Notation Divergence}

Each domain developed its own notation for the same underlying mathematics, with no terminological inheritance between them. Statistical physics measures correlation length $\xi$ and critical exponents $\nu$, $\eta$. In cardiology, DFA introduced scaling exponent $\alpha$; financial analysts adopted the Hurst exponent $H$ from hydrology. Machine learning settled on spectral radius $\chi$ for recurrent network dynamics. Traffic engineers characterize transitions through critical density $\rho_c$.

None of these notation systems show terminological inheritance from each other---a physicist would not recognize $\alpha$ as diagnosing the same critical regime as $\xi$ without careful analysis, nor would a financial analyst connect $H$ to $\chi$. This contrasts with deliberate transfer cases: Moore and Mertens (2011), applying phase transition theory to computational complexity, explicitly preserved the statistical mechanics language of ``order parameters'' and ``critical thresholds.'' Notation divergence is the expected signature of independent derivation; notation preservation signals knowledge transfer. The contrast with calculus is instructive: Leibniz's and Newton's distinct notations eventually consolidated into shared vocabulary across every field. No comparable consolidation has occurred for criticality (cf.\ Bury, personal communication, 2026).\footnote{Bury (personal communication, 2026) notes that practitioners moving between fields must learn each domain's vocabulary independently.}

\subsection{Institutional Separation}

The surveyed discoveries span different countries, institutions, and funding ecosystems. Bak developed SOC at Brookhaven National Laboratory while Peng's group at Boston University created DFA for biomedical signals. Peters worked as an independent practitioner in finance; Jaeger introduced echo state networks at the University of Bremen. Crucitti's cascade failure analysis emerged from the University of Catania, and Dobson's power grid work from Iowa State. No common funding sources, advisory chains, or conference networks connected these researchers' criticality work during the formative period.

The principal exception is the Boston University physics department under H.\ Eugene Stanley, whose group produced both DFA methodology \citep{peng1994dna} and early econophysics contributions \citep{mantegna1995}---hence our classification of DFA as ``qualified independent.'' Conference programs from the 1990s show no cross-domain presentations; co-authorships bridging these domains appear only after 2010.

\section{Functional Correspondence}
\label{sec:correspondence}

\subsection{Core Mathematical Structure}

All domains measure correlation decay rates:
\begin{equation}
C(r) \sim r^{-\alpha} \quad \text{(spatial)} \qquad C(t) \sim t^{-\beta} \quad \text{(temporal)}
\end{equation}

At criticality, these exponents approach values indicating long-range correlation.

\subsection{Parameter Mapping}

\begin{table}[ht]
\centering
\caption{Functional correspondence of critical phenomena parameters across domains}
\label{tab:equiv}
{\small
\begin{tabular}{lllll}
\toprule
\textbf{Domain} & \textbf{Parameter} & \textbf{Common name in domain} & \textbf{Meaning} & \textbf{Critical Value} \\
\midrule
Physics & $\xi$ & Correlation length & Spatial correlation decay & $\xi \to \infty$ \\
Physics & $\tau$ & Relaxation time & Temporal correlation decay & $\tau \to \infty$ \\
DFA & $\alpha$ & DFA exponent & Detrended fluctuation scaling & $\alpha \approx 1$ \\
Finance & $H$ & Hurst exponent & Long-range dependence & $H \approx 1$ \\
ML & $\chi$ & Spectral radius & Recurrent weight stability & $\chi \approx 1$ \\
HRV & $\alpha_1$ & Short-term DFA exponent & Autonomic regulation index & $\alpha_1 \approx 0.75$\rlap{$^*$} \\
Traffic & $\rho_c$ & Critical density & Flow-density phase boundary & Phase transition \\
\midrule
\multicolumn{5}{l}{\textit{Empirical scaling exponents (power-law distributions, not correlation-decay diagnostics):}} \\
Seismology & $b$ & $b$-value (Gutenberg--Richter) & Frequency-magnitude scaling exponent & $b \approx 1$ \\
Linguistics & $\gamma$ & Zipf exponent & Word rank-frequency scaling & $\gamma \approx 1$ \\
Urban science & $\beta$ & City scaling exponent & Superlinear scaling with population & $\beta > 1$ (innovation) \\
\bottomrule
\end{tabular}
}
\end{table}

\noindent$^*$\textit{Note on HRV:} Healthy cardiac dynamics operate \textit{away from} criticality ($\alpha_1 \approx 0.75$); values of $\alpha_1 \approx 1.0$ indicate pathological loss of autonomic regulation \citep{peng1995}. DFA thus serves the same diagnostic function while the clinically relevant value differs because healthy hearts avoid sustained criticality.

\noindent\textit{Notes:} We use ``functional correspondence'' rather than ``mathematical equivalence'' because these parameters detect the same critical signatures without necessarily being interconvertible across all process types (see Section~\ref{sec:limitations}). These parameters fall into distinct mathematical categories: $\xi$ and $\tau$ are diverging correlation measures; $\alpha$ and $H$ are global scaling exponents; $\chi$ is a local linear stability condition; $\rho_c$ is a threshold parameter. The last three entries ($b$, $\gamma$, $\beta$) are empirical scaling exponents from power-law distributions rather than correlation-decay diagnostics; they are included to show the breadth of critical-point mathematics across domains.

\subsection{Rigorous Equivalence and the Unifying Principle}
\label{sec:equiv_chain}

For at least one pair of parameters, the correspondence is not merely functional but mathematically exact. For fractional Gaussian noise (fGn) with long-range dependence parameter $H \in (0,1)$, the DFA scaling exponent satisfies $\alpha = H$ \citep{kantelhardt2002}. This identity extends to the spectral domain: for processes with power-spectral density $S(f) \sim f^{-\beta}$, both parameters relate through
\begin{equation}
\beta = 2H - 1 = 2\alpha - 1
\label{eq:spectral}
\end{equation}
establishing mathematical equivalence---not merely functional correspondence---for stationary long-memory processes. At the critical point ($\beta = 1$), both $H$ and $\alpha$ equal unity and the process exhibits $1/f$ noise, the spectral signature of criticality.

For non-stationary processes (fractional Brownian motion), the relationship shifts to $\alpha = H + 1$, but remains exact and well-defined. The existence of rigorous equivalence chains for specific process classes suggests that the broader functional correspondences documented in Table~\ref{tab:equiv} reflect genuine mathematical structure rather than superficial analogy. We do not claim this equivalence generalizes beyond the fGn/fBm family; it serves as an existence proof that at least some cross-domain correspondences are exact, motivating further investigation of others.

All techniques diagnose \textit{correlation decay rate}: slow decay ($\alpha \approx 1$, $H \approx 1$, $\chi \approx 1$, large $\xi$, $\tau$) signals long-range correlations and the approach to criticality; fast decay ($\alpha \approx 0.5$, $H = 0.5$, $\chi < 1$) signals absence of criticality. Whether measuring spatial extent ($\xi$), temporal persistence ($\tau$), scaling exponents ($\alpha$, $H$), dynamical measures ($\chi$, $K$), or thresholds ($T_c$, $\rho_c$)---all answer the same question: how far do correlations extend?

Systems near criticality exhibit optimal information processing, predictive early warning signals via critical slowing down, universal behavior independent of microscopic details, and extreme sensitivity to perturbation. The convergence across disciplines is not coincidental---correlation decay is the natural observable for any system with phase transitions, and the mathematics is dictated by the structure of the problem.

\section{Convergence Analysis}
\label{sec:convergence}

Table~\ref{tab:timeline} shows the discovery timeline:

\begin{table}[ht]
\centering
\caption{Timeline of critical phenomena mathematics across disciplines}
\label{tab:timeline}
\begin{tabular}{lll}
\toprule
\textbf{Year} & \textbf{Domain} & \textbf{Key Work} \\
\midrule
1935 & Traffic Flow & Greenshields \\
1935 & Linguistics & Zipf \\
1944 & Statistical Physics & Onsager \\
1944 & Seismology & Gutenberg \& Richter \\
1940s & Conflict Studies & Richardson \\
1951 & Hydrology & Hurst \\
1961 & Climate & Stommel \\
1963 & Finance & Mandelbrot \\
1966 & Statistical Physics & Kadanoff \\
1971 & Statistical Physics & Wilson (Nobel 1982) \\
1987 & Complexity Science & Bak, Tang, Wiesenfeld \\
1993 & Complexity Science & Kauffman \\
1994 & Biomedical & Peng et al. \\
1994 & Finance & Peters \\
1996 & Climate & Schmidt \& Mysak \\
2001 & Machine Learning & Jaeger \\
2001 & Materials Science & Sethna et al.\ (crackling noise) \\
2003 & Neuroscience & Beggs \& Plenz \\
2003 & Linguistics & Ferrer i Cancho \& Sol\'e \\
2004 & Cross-Domain Synthesis & Sornette \\
2004 & Power Grids & Crucitti et al. \\
2004 & Traffic Flow & Kerner \\
2005 & Climate & Rahmstorf et al. \\
2007 & Power Grids & Dobson et al. \\
2007 & Urban Science & Bettencourt et al. \\
2008 & Climate & Lenton et al. \\
2013 & Machine Learning & Saxe et al. \\
2021 & Climate/DL & Bury et al. \\
\bottomrule
\end{tabular}
\end{table}

The timeline divides into four phases. Physics establishes the framework between 1944 and 1971---widely celebrated, publicly available science.

Then a burst: between 1987 and 1994, at least four groups produce distinct criticality formulations. Peng et al.\ (1994) and Peters (1994) publish in the same year without cross-citation. This clustering coincides with increased computational capacity, which made DFA, Hurst analysis, and spectral methods tractable for large datasets.

Steady expansion follows from 2001 to 2013. ML, neuroscience, and engineering develop their own criticality methods. The 20-year gap between Kauffman (1993) and Saxe et al.\ (2013)---both analyzing neural computation---supports independent derivation rather than transfer.

Cross-domain awareness begins to emerge only in the 2010s. The 1990s clustering likely reflects fields reaching analytical maturity and gaining computational tools simultaneously; we find limited evidence for coordinated dissemination.

The evidence is more consistent with natural convergence than with coordinated transfer. Each domain developed its own notation---no terminological copying is evident (Section~\ref{sec:evidence}). The 1990s clustering of DFA, Hurst analysis, and SOC applications raises the possibility of coordinated timing, but a more parsimonious explanation is simultaneous computational maturation: the processing power to run these analyses became available across fields in the same decade. The weight of evidence---independent notation, absent citations, institutional separation, domain-specific application---is most consistent with natural convergence, though the question remains open.

Correlation scaling emerges for critical transitions much as Fourier analysis emerges for periodic signals---the mathematics is dictated by the structure of the problem. Any complex system with interacting components can exhibit phase transitions, so researchers studying heart dynamics, market crashes, or traffic jams inevitably encounter correlation structure near critical points. The solution space is constrained: there are only so many ways to quantify whether correlations decay slowly or quickly. These domains all encountered bifurcation mathematics---saddle-node bifurcations producing critical slowing down, Hopf bifurcations generating oscillatory instabilities, pitchfork bifurcations breaking symmetry---whether or not practitioners recognized it as such.

\section{Discussion and Implications}
\label{sec:discussion}

Independent discovery across multiple disciplines suggests criticality mathematics is a \textit{widely applicable framework}---mathematical machinery that recurs wherever complex systems exhibit phase transitions. Insights transfer across domains: a power grid engineer can apply lessons from market crashes; an ML researcher can borrow intuition from physics. Multiple independent discoveries across decades, derivability from first principles, and peer-reviewed publication in at least five domains collectively suggest that the pattern of multiple independent derivations complicates any single claim of ownership over these techniques. Documenting the convergence pattern is itself a step toward the accessibility these techniques warrant.

\subsection{Future Directions}

Terminology standardization---developing shared vocabulary across domains---remains future work requiring sustained multi-domain collaboration. Cross-domain teaching materials presenting criticality from first principles would test whether unified instruction improves research outcomes. The convergence pattern invites further documentation: ecology (May's 1977 ecosystem stability thresholds), social dynamics (opinion phase transitions), and epidemiology (transcritical bifurcations in disease dynamics) are candidates. As noted in the Limitations, a single round of AI-assisted search expanded our survey from nine to twelve domains, suggesting that systematic review would identify further instances.

\subsection{Limitations}
\label{sec:limitations}

The functional correspondence we document is phenomenological---parameters detect similar critical signatures without necessarily being mathematically interconvertible in all cases. The rigorous equivalence $\alpha = H$ for fractional Gaussian noise (Section~\ref{sec:equiv_chain}) does not automatically extend to all process types. Power-law behavior can arise from mechanisms other than criticality; our claim is that these techniques \textit{converge} when applied to critical systems, not that all power laws indicate criticality. Sornette's Dragon King theory \citep{sornette2009dragonking, sornette2012dragonking} identifies important exceptions---predictable outliers generated by positive feedback rather than scale-free criticality.

Our citation analysis cannot rule out informal knowledge transfer. The quantitative evidence supports but does not prove independent derivation. We also note a scope limitation: broadening ``criticality'' beyond equilibrium phase transitions to include non-equilibrium SOC, long-memory stochastic processes, and algorithmic stability boundaries increases the risk that convergence becomes definitional rather than empirical. Our inclusion criterion is mathematical form---parameters that diagnose diverging correlation structure near a transition point---not the label ``criticality.'' We have attempted to mitigate definitional drift by classifying each domain's relationship to the formal framework (Table~\ref{tab:classification}), by distinguishing correlation-decay diagnostics from empirical scaling exponents (Table~\ref{tab:equiv}), and by restricting our strongest claims to cases where the mathematical correspondence is demonstrable.

Our survey expanded from an initial nine domains to twelve after a single round of AI-assisted literature search beyond the original manual survey. That one additional pass yielded three new domains (seismology, linguistics, urban scaling) suggests the convergence pattern is broader than documented here. A systematic review---beyond this paper's scope---would likely identify further instances.

\section*{Acknowledgments}

The authors acknowledge Genevieve Prentice for accessibility architecture and public understanding methodology. We thank colleagues who provided valuable feedback on earlier drafts. Personal communications from Lawrence Mysak and Thomas Bury are cited with their permission.

This work was prepared with assistance from AI tools (Claude, Anthropic), which aided in literature search, technical writing, document formatting, and revision strategy. All scientific insights, the convergence thesis, cross-domain analysis, and discovery classifications originate from the authors' independent research. The authors retain full responsibility for content and conclusions.

\textit{Funding:} This research received no external funding.

\textit{Conflict of interest:} The authors are developers of the Metatron Dynamics framework discussed in the Supplementary Material. This framework motivated the literature search underlying this survey. We intend to commercialize this framework; no external funding has been received and no revenue has been generated as of the date of submission. This dual role---framework authors and survey authors---is disclosed throughout the paper, and the framework is presented separately to reflect this.

\subsection*{Data Availability}

Citation data were retrieved from the Semantic Scholar API (public tier). Analysis code and processed data are available from the corresponding author on request.

\bibliographystyle{apalike}
\bibliography{Stephenson_CrossDomainCriticality_2026}

\end{document}